\documentstyle[psfig,times]{mn}
\begin{document}
\hsize=6truein

\renewcommand{\thefootnote}{\fnsymbol{footnote}}

\title[]{Note on a polytropic $\beta-$model to fit the X-ray
surface brightness of clusters of galaxies}

\author[]
{\parbox[]{6.in} {Stefano Ettori \\
\footnotesize
Institute of Astronomy, Madingley Road, Cambridge CB3 0HA \\
settori@ast.cam.ac.uk
}}                                            
\date{{\bf MNRAS, 311, 313 (2000)}}
\maketitle

\begin{abstract}
In this note, I suggest that the $\beta-$model used to fit the
X-ray surface brightness profiles of extended sources, like
groups and clusters of galaxies, has to 
be corrected when the counts are collected in a wide energy band
comparable to the mean temperature of the source and 
a significant gradient in the gas temperature is observed.
I present a revised version of the $\beta-$model for the X-ray
brightness that applies to a
intracluster gas with temperature and density related by a polytropic
equation and extends the standard version that is strictly valid for 
an isothermal gas.
Given a temperature gradient observed through an energy window of 
1--10 keV typical for the new generation of X-ray observatories, 
the $\beta$ parameter can change systematically up to 20 per cent 
from the value obtained under isothermal assumption, 
i.e. by an amount larger that any 
statistical uncertainty obtained from the present data. 
Within the virial regions of typical clusters of galaxies, these
systematic corrections affect the total gravitating mass estimate
by 5--10 per cent, the gas mass by 10--30 per cent and
the gas fraction value up to 50 per cent, when compared 
to the measurements obtained under the isothermal assumption.
\end{abstract}


\section{INTRODUCTION} 

The $\beta-$model is widely used in the X-ray astronomy to parametrise
the gas density profile in groups and clusters of galaxies fitting
their surface brightness profile. Cavaliere \& Fusco-Femiano (1976, 1978)
note that `` ... since both gas and galaxies
distributions conforms to the same gravitational potential, the former can
be {\it directly} related to the latter; the galaxies may be considered as
tracers of the total potential well ...":
\begin{equation}
\frac{1}{\rho_{\rm gal}} \frac{d P_{\rm gal}}{dr} =
\frac{1}{\rho_{\rm gas}} \frac{d P_{\rm gas}}{dr} 
\end{equation}
Using the King approximation (1962) to the inner portions of an 
isothermal sphere (Lane-Emden equation in Binney \& 
Tremaine 1987; note that the King approximation is proportional to 
$r^{-3}$ at the outer radii, whereas 
the isothermal sphere is proportional to $r^{-2}$; cf. Figure~1
\footnote[1]{Throughout this note, a Hubble constant of 50 $h_{50}$ km s$^{-1}$
Mpc$^{-1}$ is considered.}):
\begin{equation}
\rho_{\rm gal} = \rho_{\rm 0, gal} (1+x^2)^{-3/2}, \ x=r/r_{\rm c} 
\end{equation}
and the perfect gas law, one obtains the formula:
\begin{equation}
\rho_{\rm gas} = \rho_{\rm 0,gas} (1+x^2)^{-3 \beta /2},
\label{eq:rho_b}
\end{equation} 
where the different energy distribution of the gas and galaxies 
is parameterized by using the parameter 
$\beta \sim (\sigma^2_{\rm gal}/T_{\rm 
gas})$, where $\sigma_{\rm gal}$ is the galaxies velocity dispersion and
$T_{\rm gas}$ is the temperature of the gas.

\begin{figure}
\psfig{figure=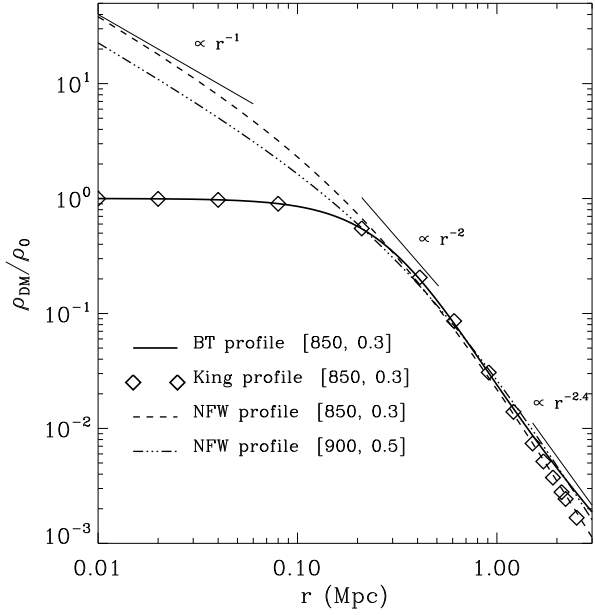,width=.5\textwidth}
\caption[Dark matter density profiles]
{The Binney \& Tremaine (BT, from equation 4-125 in Binney \&
Tremaine 1987) dark matter profile for the self-gravitating 
isothermal sphere is here compared for different input parameters,
[$\sigma$ (km s$^{-1}$), $r_{\rm c}$ (or $r_{\rm s}$, Mpc)], 
to the Navarro-Frenk-White (NFW, 1995) profile that comes from
extended and highly resolved numerical simulations of clusters
of galaxies. 
Both of these are also compared with the King approximation (King 1962)
to the inner part of the self-gravitating isothermal sphere.
All of them are normalized to the central value of the self-gravitating 
isothermal
sphere profile [$\rho_0 = 9 \sigma^2 / (4 \pi G r_{\rm c}^2) = 9.05
\times 10^{-26}$ g cm$^{-3}$]. Inside the core radius, the NFW profile
does not flatten like the BT profile.
In the outer part of the region of interest (above 1 Mpc), agreement 
between the two profiles is obtained by increasing the
velocity dispersion and the core radius (or {\it scale radius})
in the NFW profile. Fitting a power law, it can be shown that around
$2.5 \times r_{\rm s} \sim 2$ Mpc  
the NFW profile approaches a $r^{-2.4}$ form.
} \label{NFWdens} \end{figure}

The surface brightness profile observed at the projected radius $b$,
$S(b)$, is the projection on the sky of the plasma emissivity,
$\epsilon (r)$:
\begin{equation}
S(b) \equiv S_{\rm b} = \int_{b^2}^{\infty} \frac{\epsilon \ dr^2}
{\sqrt{r^2 - b^2}}.
\label{eq:sb} \end{equation}
(Hereafter I adopt $r$ as symbol for the projected
radius $b$).

The emissivity is equal to
\begin{equation}
\epsilon (r) = \Lambda(T_{\rm gas}) \ n_{\rm p}^2
\ {\rm erg} \ {\rm s}^{-1} \ {\rm cm}^{-3},
\label{eq:em} \end{equation}
where $n_{\rm p} = \rho_{\rm gas} / (2.21 \mu m_{\rm p} )$ is the
proton density and the cooling function, $\Lambda(T_{\rm gas})$,
depends upon the mechanism of the emission and can be represented as
\begin{equation}
\Lambda(T_{\rm gas}) = \overline{g} \ \lambda' \ T_{\rm gas}^{\alpha'}
= \lambda \ T_{\rm gas}^{\alpha},
\label{eq:cool} \end{equation}
where $\overline{g}$ is the velocity averaged Gaunt factor that is
equal to about 1.2 within an accuracy of 20\% for a bolometric
emissivity (for example, at $T_{\rm gas} >$ 2.5 keV, the emission
is mainly due to bremsstrahlung and the cooling function can be written
with $\lambda \sim 10^{-23}$ and $\alpha = 0.5$; 
see, e.g., Sarazin 1988).

Assuming isothermality and a $\beta$-model for the gas density 
(eq.~\ref{eq:rho_b}),
the surface brightness profile has an analytic solution:
\begin{eqnarray}
S_{\rm b} & = & \sqrt{\pi} n_0^2 r_{\rm c} \Lambda(T_{\rm gas}) 
\frac{\Gamma (3\beta -0.5)}{\Gamma (3\beta) } (1 + x^2)^{0.5 -3 \beta} 
\nonumber \\
 & = & S_0 (1+x^2)^{0.5 -3 \beta},
\label{eq:beta}
\end{eqnarray}
that is strictly valid under the condition that $3 \beta > 0.5$ and
the cooling function $\Lambda(T_{\rm gas})$ does not change radially.

In this note I will focus on energies typical for clusters of galaxies, 
considering the fact that there is now evidence for a decrease
in the gas temperature in the outer parts of clusters
(Markevitch et al. 1998).
Even if these results conflict with studies from other groups
which indicate that clusters are generally isothermal 
(e.g. Irwin et al. 1999, Kikuchi et al. 1999, White 1999), I
highlight how a temperature gradient can affect the
estimate of the $\beta$ parameter.

In the following discussion, I will assume that the cluster gas 
density, $n_{\rm gas}$, is well described by 
a $\beta-$ model and the gas is in the polytropic state, so that
\begin{equation}
\frac{T_{\rm gas}}{T_0}=\left( \frac{n_{\rm gas}}{n_0} \right)^{\gamma -1},
\label{eq:poly}
\end{equation}
where the polytropic index $\gamma$ ranges between 1 and 5/3, the limits
corresponding to the gas being isothermal and adiabatic, respectively.

\section{A polytropic $\beta-$model}

\begin{figure}
\psfig{figure=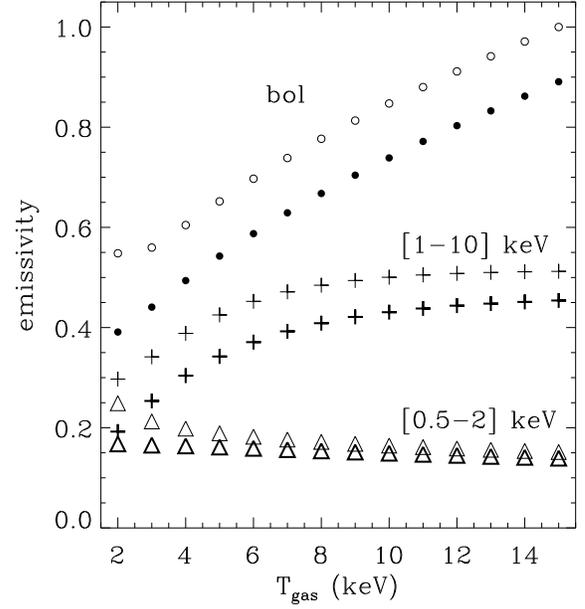,width=.5\textwidth}
\caption[]
{This plot shows the total dependence of the emissivity upon the 
temperature in different energy window [{\it bolometric}, 1-10 keV, 
0.5-2 keV] and for a cluster whose temperature ranges between 2 and 
15 keV (where the bremsstrahlung emission dominates).
The curves are calculated using a MEKAL model (Kaastra 1992,  Liedahl
et al. 1995) in XSPEC (version~10, Arnaud 1996)
for two different values of metallicity: 0.3 (thickest symbols) and 1.0
times the solar abundance. They are normalized to the bolometric
emissivity at $T_{\rm gas} = 15$ keV with $Z = 1 Z_{\odot}$.
When the free-free radiation dominates, these curves can be calculated
analytically   integrating over the window energy [$E1-E2$] the Gaunt 
factor, assumed equal to $0.9 (E/kT)^{-0.3}$, multiplied by
$(kT)^{-0.5} e^{-(E/kT)}$.
Note that the emissivity is almost independent from the temperature 
of the plasma in the energy accessible to the {\it ROSAT} observatory
(cf. Table~1).
} \end{figure}

\begin{table}
\caption[]{For a given energy window $\Delta E$, 
I calculate the slope, $\alpha$, of the power law in eqn.~\ref{eq:cool} 
over a selected temperature range $\Delta T_{\rm gas}$. $\lambda$ is 
fixed to the value of the emissivity corresponding to the higher
$T_{\rm gas}$, i.e. $\alpha = \log(\epsilon /\epsilon^{\rm max}) /
\log(T_{\rm gas}/T_{\rm gas}^{\rm max})$.
}
\begin{tabular}{cccc}
$\Delta E$ (keV) & $\Delta T_{\rm gas}$ (keV) & $\alpha$ 
$(Z=0.3 Z_{\odot})$ & $\alpha$ $(Z=1 Z_{\odot})$ \\
 & & & \\
bol    & 5-10 & 0.45 & 0.39 \\
    & 5-7 & 0.44 & 0.37 \\
    & 3-6 & 0.43 & 0.36 \\
    & 3-5 & 0.42 & 0.34 \\
0.5-2  & 5-10 & --0.13 & --0.20 \\
    & 5-7 & --0.10 & --0.20 \\
    & 3-6 & --0.08 & --0.21 \\
    & 3-5 & --0.06 & --0.22 \\
1-10  & 5-10 & 0.25 & 0.16 \\
    & 5-7 & 0.37 & 0.27 \\
    & 3-6 & 0.47 & 0.36 \\
    & 3-5 & 0.54 & 0.41 \\
\end{tabular}
\end{table}

When the cluster temperature is above the energy range of the detector
with a narrow bandpass (e.g. ROSAT), the emission measure will have 
a negligible dependence on the temperature (Figure~2).

However, this dependence becomes significant when the energy band is 
wide and its mean energy range is comparable to the mean temperature 
of the cluster.
For example, the emissivity due to free-free radiation for a plasma 
temperature of 10 keV is 50 per cent larger than one 
at 4 keV in the energy range [1-10] keV, whereas it changes
by 1 per cent in the ROSAT bandpass [0.5-2] keV.
In Figure~2, I show how the total cluster emissivity convolved with 
a given energy bandpass of a X-ray telescope depends upon the plasma
temperature. 

A proper deprojection analysis (cf. White et al. 1997, Ettori \& 
Fabian 1999) will be necessary to evaluate the emissivity 
in differential volume shells and recover both the gas
density and temperature profile. On the other hand, this technique
is computationally very expensive and a simple fitting procedure
can be preferred in most cases where the gas and total mass 
distributions are investigated.

As discussed above, Cavaliere \& Fusco-Femiano have introduced
the $\beta-$model for an isothermal gas distribution.
But if, as usually done, the gas density distribution is obtained
from the parameters of the best fit of eqn.~\ref{eq:beta} 
for the surface brightness profile, any temperature 
gradient present in the plasma will not be taken properly into account.
To consider this correction, I assume that the gas density profile
is well described by a $\beta-$model (eqn.~\ref{eq:rho_b}) and use
the temperature dependence of the emissivity (given in eqn.
~\ref{eq:cool} through the parameter $\alpha$) and a polytropic 
relation between gas density and temperature (represented by the index
$\gamma$ in eqn.~\ref{eq:poly}) in the definition of the 
surface brightness (eqn.~\ref{eq:sb}).
Then, eqn.~\ref{eq:beta} can be re-written as
\begin{eqnarray}
S_{\rm b} & = & \int_{b^2}^{\infty} \lambda \ T^{\alpha}_{\rm gas}
\ n^2_{\rm gas} \ dr^2/\sqrt{r^2 - b^2} \nonumber \\
 & = & \int_{b^2}^{\infty} n^2_0 \lambda \ (1+x^2)^{-3 \beta 
\left(1 + \alpha \frac{\gamma -1}{2} \right) } \nonumber \\
 & = & \sqrt{\pi} n_0^2 r_{\rm c} \lambda \frac{\Gamma (3 \beta'
-0.5 )}{\Gamma (3\beta')} (1+x^2)^{0.5 -3 \beta'} \nonumber \\  
 & = & S_0 \left[1 + \left(\frac{r}{r_{\rm c}} \right)^2 
\right]^{0.5 -3 \beta \left(1 + \alpha \frac{\gamma -1}{2} \right) }.
\label{eq:b_poly}
\end{eqnarray}
Here $\beta' = \beta \left(1 + \alpha \frac{\gamma -1}{2} \right)$ 
represents the uncorrected measured value.

In Table~1, I calculate the $\alpha$ values for a set of interesting
cases.

It is worth noting that, even if the observed gas temperature, $T_{\rm gas}$,
is the projection on the sky of the real temperature, 
$T^{\rm real}_{\rm gas}$ (\ref{eq:poly}), 
weighted by the cluster emission, $T_{\rm gas}$ 
does not depend upon the parameter $\alpha$ (see also Markevitch et al. 1999):
\begin{eqnarray}
T_{\rm gas}^{\rm proj}(b) & \equiv & T_{\rm gas} = 
\frac{ \int_{b^2}^{\infty} \epsilon \ T^{\rm real}_{\rm gas} \ 
dr^2/\sqrt{r^2 - b^2} }{\int_{b^2}^{\infty} \epsilon
\ dr^2/\sqrt{r^2 - b^2} } \nonumber \\
 & \propto & \frac{ (1+x^2)^{0.5-1.5\beta
[2+(\gamma -1)(\alpha +1)]} }{(1+x^2)^{0.5-1.5\beta [2+\alpha (\gamma
-1)]} } \nonumber \\
 & \propto & (1+x^2)^{-1.5\beta (\gamma -1)}.
\label{eq:tproj}
\end{eqnarray}

I can then estimate the corrections that a polytropic temperature
profile produces on the uncorrected values ($\beta ', 
\gamma '$). To do this, I solve the system of equations
given by the surface brightness and temperature profiles modeled with
a $\beta-$model:
\begin{equation}
\left\{ \begin{array}{l}
S_0 (1+x^2)^{0.5-3 \beta '} = S_0 (1+x^2)^{0.5-3 \beta 
\left(1 + \alpha \frac{\gamma -1}{2} \right) } \\
T_0 (1+x^2)^{-1.5 \beta ' (\gamma ' -1)} = T_0 (1+x^2)^{-1.5 \beta 
(\gamma -1)}
\end{array}
\right.
\end{equation}
that can be simplified to:
\begin{equation}
\left\{ \begin{array}{l}
\beta ' = \beta \ \left(1+ \alpha \frac{\gamma-1}{2} \right) \\
\beta ' \ (\gamma ' -1) = \beta \ (\gamma -1) 
\end{array}
\right.
\end{equation}
This system has the following solution:
\begin{equation}
\left\{ \begin{array}{l}
\beta = \beta ' \ (1 +\alpha/2 -\alpha \gamma '/2)  \\
\gamma = \frac{\gamma ' + \alpha/2 - \alpha \gamma '/2}{1+\alpha/2
-\alpha \gamma '/2} 
\end{array}
\right.
\end{equation}

Figure~3 shows the relative systematic corrections that affect
the values of $(\beta, \gamma)$ for a given set of 
($\alpha, \gamma'$).

\begin{figure}
\psfig{figure=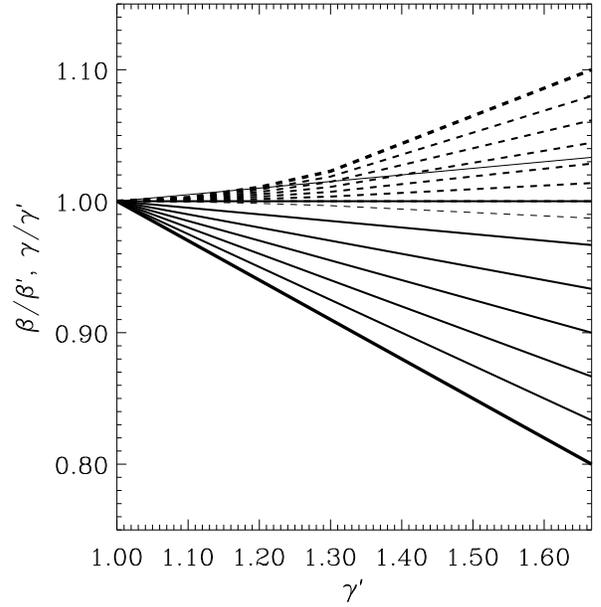,width=.5\textwidth}
\caption{This plot shows the corrections of the parameters
$\beta$ (solid line) and $\gamma$ (dashed line) for a given temperature
dependence of the emissivity, $\alpha$, and an uncorrected
polytropic index $\gamma'$. The thickest lines correspond to 
the case $\alpha=0.6$; the thinest lines are for $\alpha=-0.1$.
The other lines spans $\alpha$ values between 0 and 0.5.
} \end{figure}

In practice, once a temperature profile is measured in a known energy
bandpass, the $\alpha$ parameter can be defined. 
The conversion from the count rate to the flux can be done assuming
the central, highest temperature (i.e. with the largest emissivity,
cf. Table~1).
The functional forms, $S_{\rm b}=S_0 (1+x^2)^a$ and $T_{\rm gas} =
T_0 (1+x^2)^b$, can be then fitted to the surface brightness and 
temperature profiles, respectively.
If they still represent a good model of the data, the
correct $(\beta, \gamma)$ values can be estimated:
\begin{equation}
\left\{ \begin{array}{l}
\beta = \frac{0.5 -a -\alpha b}{3}  \\
\gamma = 1 + \frac{2b}{0.5 -a -\alpha b}
\end{array}
\right.
\end{equation}

The consequences of these corrections are discussed below.

\section{DISCUSSION AND CONCLUSIONS}
I suggest that the $\beta-$model used to fit the 
X-ray surface brightness
profiles of clusters of galaxies has to be corrected when the
data from the next generation of X-ray observatories will be 
available. In fact, to include any temperature gradient 
that will affect the cluster emissivity as observed in a large
energy window, we have to extend the use
of the $\beta-$model to the polytropic case. 
The new form of this {\it polytropic} $\beta-$model is given in 
eqn.~\ref{eq:b_poly}.

The spatially resolved surface brightness profiles obtained 
with the ROSAT observatory allow 
estimates of $\beta$ with an accuracy ($1 \sigma$) of about
2 per cent (e.g., Mohr et al. 1999, Neumann \& Arnaud 1999). 
However, the limited energy band pass of ROSAT
does not arise any problem on the application of the $\beta-$model 
when a temperature gradient is observed in the intracluster
medium. This is also true for any surface brightness profile
that is obtained collecting photons within an instrumental band pass 
at energies
lower than the mean plasma temperature. In particular, considering
that the effective area of the present X-ray detectors is larger
at $E \sim 1-2$ keV,
an energy window around these values can be a good choice 
for a temperature-independent emissivity for hot clusters of galaxies.
But an energy range around few keV is still problematic 
for cool clusters and groups of galaxies that present temperature
gradients.

\begin{table}
\caption[]{Changes of the derived quantities after that the 
corrections of the best-fit parameters are considered:
${\rm unc}(\%) = 100 \times (Q_{\rm corr} - Q_{\rm no-corr})/Q_{\rm no-corr}$.
}
\begin{tabular}{cccc}
 $r$  & $\Delta M_{\rm gas}$ & $\Delta M_{\rm tot}$ & $\Delta f_{\rm gas}$ \\
 & & & \\
\multicolumn{4}{c}{$\alpha =0.47, \ \beta =2/3, \ \gamma =1.20$} \\
1 Mpc     &  $+8$\%  & $-4$\% & $+12$\% \\
2 Mpc     &  $+13$\% & $-4$\% & $+18$\%  \\
$r_{200}$ &  $+11$\% & $-5$\% & $+17$\% \\
 & & & \\
\multicolumn{4}{c}{$\alpha =0.50, \ \beta =2/3, \ \gamma =5/3$} \\
1 Mpc     &  $+31$\% & $-10$\% & $+46$\% \\
2 Mpc     &  $+56$\% & $-10$\% & $+74$\%  \\
$r_{200}$ &  $+34$\% & $-10$\% & $+48$\%  \\
\end{tabular}
\end{table}

The new generation of X-ray observatories,
e.g., Chandra\footnote[2]{{\tt http://asc.harvard.edu/}} and 
XMM\footnote[3]{ {\tt http://astro.estec.esa.nl/XMM/} }, 
will operate in a wider energy
band (e.g. [1-10] keV) than ROSAT and will provide a more accurate
estimates of the parameters of the $\beta-$model.
Therefore, as shown above, the presence of a plasma temperature gradient
will affect the use of the $\beta-$model with a systematic 
uncertainty comparable or larger than any statistical error. 
In particular, the estimate of (i) the total gravitating
mass, $M_{\rm tot} \propto \beta \ \gamma \ (1+x^2)^{-1.5 \beta (\gamma -1)}$ 
(cf. eqn.~\ref{app:mtot}), (ii) the gas mass, 
$M_{\rm gas} \propto \int (1+x^2)^{-1.5 \beta} \ x^2 dx$, 
and (iii) the consequent gas fraction, $f_{\rm gas}$, 
can be quite significantly affected.
For example, given a cluster with a typical core radius, $r_{\rm c}$, 
of 0.3 Mpc and a radial decrease of the plasma temperature
from 6 to 3 keV with uncorrected parameters $\beta' = 2/3$ and $\gamma' = 1.20$
(e.g. Markevitch et al. 1999), corrections
of $+8, -4, +12$ per cent on $M_{\rm gas}, M_{\rm tot}, f_{\rm gas}$,
respectively, will be necessary at $r = 1$ Mpc.
At the more physically meaningful radius $r_{200}$, 
where the mean cluster density is $\Delta = 200$ times the critical value
(cf. eqn.~\ref{app:r500} in Appendix),
the corrections are $+11, -5, +17$ per cent, respectively. 
These corrections increase considerably up to 
$+56, -10, +74$ per cent at $r = 2$ Mpc 
for $M_{\rm gas}, M_{\rm tot}, f_{\rm gas}$, respectively, when 
$\alpha=0.5$, $\beta'=2/3$, $\gamma'=5/3$ (cf. Table~2).

\section*{ACKNOWLEDGEMENTS} I acknowledge the support
of the Royal Society. Andy Fabian and David White are thanked 
for an useful reading of the manuscript.

\appendix
\section{Other analytic formulae}
I write here the analytic expressions for the derived quantities like
the total gravitating mass, $M_{\rm tot}$, and the radius at which the 
mean density within a cluster at redhsift $z$ is $\Delta$ times the background,
$r_{\Delta}$, when the gas density is assumed well described by 
a $\beta-$ model and presents a polytropic ($1 \le \gamma \le 5/3$) 
dependence upon the gas temperature:  

\begin{eqnarray}
M_{\rm tot} (r) & = & -\frac{T(r) \ r}{G \mu m_{\rm p}} \left( 
\frac{\partial \ln \rho}{\partial \ln r} + \frac{\partial \ln T}{\partial 
\ln r}\right)  \nonumber \\
 & = & \frac{3 \ \beta \gamma \ T_0 \ r_{\rm c}}{G \mu m_{\rm p}}
\frac{x^3}{ (1+x^2)^{B} } \nonumber \\
 & = & \frac{1.060 \times 10^{14}}{\mu} \beta \gamma T_0 r_{\rm c} 
\frac{x^3}{ (1+x^2)^{B} } \ M_{\odot},
\label{app:mtot}
\end{eqnarray}

[Henriksen \& Mushotzky 1986; see also Cowie, Henriksen \& Mushotzky 1987, 
Hughes et al. 1988 for a critical discussion on the presence of an artificial
cutoff in the parameter space, i.e. $2B < 3$ or $\gamma < 1 +1/(3\beta)$, 
to avoid the related virial density falling to zero (or negative values) 
at small radii]
and, given that $M_{\rm tot} (r_{\Delta}) = \frac{4}{3} \pi \rho_{\rm c}
(1+z)^3 r_{\Delta}^3 \Delta$, 

\begin{eqnarray}
\frac{r_{\Delta}}{r_{\rm c}} & = & 
\sqrt{ \left( \frac{ 3 \ \beta \gamma \ T_0}
{G \mu m_{\rm p} (4/3) \pi \rho_{\rm c} (1+z)^3 r_{\rm c}^2 \Delta } 
\right)^{\frac{1}{B}} -1 } \nonumber \\
 & = & \sqrt{ \left( \frac{229.5}{\mu (1+z)^3} 
\frac{\beta \gamma \ T_0}{r_{\rm c}^2 \Delta} \right)^
{\frac{1}{B}} -1 },
\label{app:r500}
\end{eqnarray}
where the exponent $B = 1.5 \beta (\gamma -1) +1$, 
$T_0$ is the central temperature in keV, $r_{\rm c}$ the core radius
in $h_{50}^{-1}$ Mpc, 
$\mu$ is the mean molecular weight in a.m.u. and the numerical values 
include the gravitational constant $G$, the mass of the proton
$m_{\rm p}$ and all the unit conversions.

\end{document}